\documentstyle[prb,aps,psfig]{revtex}
\def\be{\begin{equation}}
\def\ee{\end{equation}}
\def\bc{\begin{center}}
\def\ec{\end{center}}
\newcommand{\D}{{\rm d}}

\begin{document}

\input epsf.sty
\twocolumn[\hsize\textwidth\columnwidth\hsize\csname %
@twocolumnfalse\endcsname
 
\draft

\widetext
\title{Anisotropic scaling and generalized conformal invariance 
at Lifshitz points}

\author{Michel Pleimling$^{a,b,*}$ and Malte Henkel$^a$ }

\address{
$^a$ Laboratoire de Physique des Mat\'eriaux,$^{**}$
Universit\'e Henri Poincar\'e Nancy I, B.P. 239, \\
F -- 54506 Vand{\oe}uvre l\`es Nancy Cedex, France\\
$^b$ Institut f\"ur Theoretische Physik I, Universit\"at Erlangen-N\"urnberg,
D -- 91058 Erlangen, Germany}
\maketitle
 
\begin{abstract}
The behaviour of the $3D$ axial next-nearest
neighbour Ising (ANNNI) model at the uniaxial Lifshitz point is studied
using Monte Carlo techniques. A new variant of the Wolff cluster algorithm 
permits the analysis of systems far larger than in previous studies.
The Lifshitz point critical exponents are $\alpha=0.18(2)$, $\beta=0.238(5)$
and $\gamma=1.36(3)$. Data for the spin-spin correlation
function are shown to be consistent with the explicit scaling function 
derived from the assumption of local scale invariance, 
which is a generalization of conformal
invariance to the anisotropic scaling {\em at} the Lifshitz point.
\end{abstract}


\phantom{.}
]

\narrowtext

The modern understanding of critical phenomena is governed by 
the notion of scale invariance.\cite{Car96} For isotropic critical
systems the extension from global, spatially homogeneous scaling
to space-dependent rescaling factors leads to the requirement of conformal
invariance of the correlators. This approach has proven
to be very fruitful when examining isotropic equilibrium critical
systems, especially in two dimensions.\cite{Car96,Bel84,Hen99}

On the other hand, little is known for systems with strongly anisotropic
critical points, where the value of the anisotropy exponent
$\theta$ differs from unity. In these cases, the two-point function 
$C( \vec{r}_{\perp}, r_{\|} )$ satisfies the scaling form
\begin{equation} \label{Gl:DynS}
C( \vec{r}_{\perp}, r_{\|} ) = b^{-2x} C( b \vec{r}_{\perp}, b^{\theta} r_{\|})
= r_{\perp}^{-2x}\, \Omega( r_{\|} r_{\perp}^{-\theta})
\end{equation} 
where $r_{\|}$ and $r_{\perp}= | \vec{r}_{\perp} |$ 
are the distances parallel and 
perpendicular with respect to a chosen axis, $x$ is a 
scaling dimension, $\theta$ is the
anisotropy exponent and $\Omega(v)$ is a scaling function. Scale invariance
alone is not enough to determine the form of the function $\Omega(v)$. 

Recently, a generalization of conformal invariance involving local
space-time-dependent scale transformations for anisotropy exponents
$\theta \neq 1$ has been proposed.\cite{Hen97} This approach attempts to
generalize the scaling (\ref{Gl:DynS}), usually considered with $b$ constant,
to space-dependent rescaling $b=b(\vec{r}_{\perp},r_{\|})$, 
thereby assuming that the two-point functions still transform in a simple way. 
These transformations are constructed, starting from the conformal 
transformations $r_{\|}\to (\alpha r_{\|}+\beta)/(\gamma r_{\|}+\delta)$ with 
$\alpha\delta-\beta\gamma=1$, in such a way that the transformations in the
`spatial' coordinates $\vec{r}_{\perp}$ are consistent with the anisotropic
scaling (\ref{Gl:DynS}). Systems which are invariant 
under these transformations and whose correlators, 
generalizing (\ref{Gl:DynS}), transform covariantly under them, are said to
satisfy {\em local scale invariance} (LSI). 
It turns out that if $\theta=2/N$, where $N$ is a positive integer, 
$\Omega(v)$ satisfies the differential equation\cite{Hen97}
\begin{equation} \label{Gl:DiffGl}
\alpha_1 \frac{\D^{N-1} \Omega(v)}{\D v^{N-1}} - v^2 \frac{\D \Omega(v)}{\D v} 
-\zeta v \Omega(v) = 0
\end{equation}
where $\zeta=2x/\theta$ and $\alpha_1$ is a constant. 
Eq. (\ref{Gl:DiffGl}) can be explictly solved\cite{Hen97} in terms of 
hypergeometric functions ${{_2}F_{N-1}}$ (conformal invariance is 
reproduced\cite{Hen97} for $N=2$ and $N=1$ gives the 
non-relativistic Schr\"odinger invariance\cite{Hen93}). 
Evidently, the above hypothesis of LSI
in systems satisfying (\ref{Gl:DynS})
is a strong one and relies on certain assumptions about the structure of the
underlying field theory. 
In this letter, we shall test the idea of LSI by explicitly
checking the resulting expressions for $\Omega(v)$ in a non-trivial spin system
which satisfies the strongly anisotropic scaling (\ref{Gl:DynS}). 

While dynamical scaling (\ref{Gl:DynS}) occurs in critical dynamics (then
$\theta$ is referred to as dynamical exponent)
or in true non-equilibrium phase transitions such as directed percolation, 
well-known examples of stongly anisotropic {\em equilibrium} criticality are 
the {\em Lifshitz points\ }\cite{Hor75}  encountered in systems
with competing interactions. At a Lifshitz point, a disordered,
an uniformly ordered and a periodically ordered phase
become indistinguishable.\cite{Hor75} 
The simplest model for these is the ANNNI
(axial next-nearest neighbour Ising) model\cite{Sel92} which 
describes faithfully, among others, magnetic systems, alloys or 
uniaxially modulated ferroelectrics.\cite{Sel88,Yeo88,Neu98}
Recently, a large variety of new physical systems (ferroelectric liquid 
crystals,\cite{Ska00} uniaxial ferrolectrics,\cite{Vys92} 
block copolymers\cite{Bat95} or even quantum systems\cite{Sch98}) 
were shown to possess a Lifshitz point which has stimulated renewed
interest in its properties. Furthermore, 
new field theory studies\cite{Mer98,Die00,Die01,Lei00} have
lead to more refined estimates (in the framework of an $\epsilon$-expansion)
of the critical exponents of the general $m$-fold Lifshitz points
in $d$ dimensions with a $n$-component order 
parameter.\cite{Mer98,Die00,Die01}

Here, we study the $3D$ ANNNI model, defined on a cubic lattice
with periodic boundary conditions. 
The Hamiltonian is 
\begin{eqnarray}
{\cal H} &=& - J \sum\limits_{xyz} s_{xyz} \left( s_{(x+1)yz} + s_{x(y+1)z} +
s_{xy(z+1)} \right) \nonumber \\
& &+ \kappa \, J \sum\limits_{xyz} s_{xyz}s_{xy(z+2)} \label{Gl:annni}
\end{eqnarray}
with $s_{xyz} = \pm 1$, whereas $J >0$ and $\kappa > 0$ are coupling
constants. In $z$-direction competition between ferromagnetic
nearest neighbour and antiferromagnetic next-nearest neighbour couplings
takes place, leading to a rich phase diagram with a multitude of
different phases, both commensurate and incommensurate to the
underlying cubic lattice.\cite{Sel92} The anisotropy exponent 
$\theta=\nu_{\|}/\nu_{\perp}$, 
where $\nu_\parallel$ und $\nu_\perp$ are the exponents 
of the correlation lengths 
parallel and perpendicular to the $z$-axis. At the uniaxial Lifshitz point,  
a recent careful field-theoretical calculation\cite{Die00} shows that 
$\theta = \frac{1}{2} - a \epsilon^2 + O(\epsilon^3)$ 
in a second-order $\epsilon$-expansion (with $\epsilon=4.5-d$)
where $a\simeq 0.0054$ for the $3D$ ANNNI model.

Our main purpose will be the numerical computation and thorough analysis of
the critical spin-spin correlation function {\em at} the uniaxial Lifshitz 
point of the ANNNI model through a large-scale Monte Carlo simulation. 
The agreement between our numerical results and the analytic expression for 
$\Omega(v)$ derived from (\ref{Gl:DiffGl}) presents evidence that local
scale invariance, as formulated in Ref. \onlinecite{Hen97}, is realized as a 
new symmetry in strongly anisotropic equilibrium critical systems.

Such a study does require reliable and precise estimates of the critical
exponents. However, published estimates of critical exponents obtained with
different techniques spread considerably, see Table I. 
We therefore undertook large-scale Monte Carlo
simulations to estimate the exponents reliably. Previous Monte Carlo 
studies\cite{Kas85} considered only small systems of (mostly) cubic 
shape. Here, we present calculations for large systems of 
anisotropic shape with $L \times L
\times N$ spins, with $20\leq L\leq 240$ and $10\leq N\leq 100$, 
taking into account the special finite-size effects coming
from the anisotropic scaling at the Lifshitz point.\cite{Bin89} 
This is the first study of the ANNNI model where the exponents 
$\alpha$, $\beta$, and $\gamma$ are computed independently.

As usual, the problems coming from critical slowing-down encountered when 
using local Monte Carlo dynamics, are alleviated by using non-local methods,
such as the Wolff cluster algorithm.\cite{Wol89}
For the Ising model with only a nearest neighbour coupling $J$,
this algorithm may be described as follows: one chooses randomly
a lattice site, the seed, and then builds
up iteratively a cluster by including a lattice site $j$ (with spin $s_j$),
neighbour to a cluster site $i$ (with spin $s_i$),
with probability $p= \frac{1}{2} \left( 1 + \mbox{sign} \left(
s_i s_j \right) \right) \left( 1 - \exp \left[ - 2 J /(k_B T) \right] \right)$.
One ends up with a cluster of spins having all the same sign which is then
flipped as a whole. This kind of same-sign clusters are 
obviously not adapted to our problem because of the competing interactions 
along the z-direction, see (\ref{Gl:annni}).

We therefore propose the following modified cluster algorithm.
Starting with a randomly chosen seed, one again builds up iteratively
a cluster. Consider a newly added cluster lattice site $i$ with spin $s_i$.
A lattice site $j$ with spin $s_j$ nearest neighbour to $i$ is included
with probability $p_{\rm n}=p$, whereas an {\it axial} next-nearest
neighbour site $k$ with spin $s_k$ is included with probability
$p_{\rm a}=\frac{1}{2} \left( 1 - \mbox{sign} \left(
s_i s_k \right) \right) \left( 1 - \exp \left[ -  2 J \kappa 
/(k_B T) \right] \right)$. Thus, the final cluster, which will be flipped
as a whole, contains spins of both signs. Ergodicity and detailed
balance are proven as usual. This algorithm works extremely
well in the ferromagnetic phase and in the vicinity of the Lifshitz
point. Generalization to other systems with competing interactions
is straightforward. For the computation of the spin-spin correlation 
function we adapt in a similar way a recently proposed very efficient 
algorithm using Wolff clusters.\cite{Eve00}
This algorithm yields the infinite-system correlation functions at 
temperatures above $T_c$ and largely reduces finite-size effects 
at the critical temperature as compared to a more traditional approach.
These algorithms will be discussed in detail elsewhere.\cite{Ple00}

We now outline the determination of the critical exponents. The results
are listed in Table I. As an example, Figure 1 shows the effective exponent
\begin{equation}
\beta_{\rm eff} = \frac{\D \ln m }{\D \ln t},
\end{equation}
where $m$ denotes the magnetization and $t = 1 - T/T_c$.
In the limit $t \to 0$ the effective exponent yields
the critical exponent\cite{Ple98} $\beta$, provided finite-size 
effects can be neglected. The two sets of data in Figure 1
correspond to two different paths in the temperature-interaction
space, both ending at the point ($\kappa=0.270$, $T_c=3.7475$), 
setting $J / k_B=1$.
For set (a) $\kappa$ was fixed at 0.270, whereas for set (b)
$\kappa= 0.270 + 1.6 \, (1/T - 1/T_c )$. The corrections to scaling
for set (b) are small compared to set (a), resulting in a plateau
for $t \leq 0.06$, thus making a very precise estimation of $\beta$ possible.
Of course, finite-size effects have to be monitored carefully. As usual,
we adjust the system size in order to circumvent
finite-size dependences.\cite{Ple98,Ple00}
For the determination of the susceptibility and specific heat critical 
exponents $\gamma$ and $\alpha$, see Table I, data obtained at temperatures
both below and above $T_c$ were analysed. Our error bars take into account
the sample averaging as well as the uncertainty in the location of the 
Lifshitz point. Based on our data, we locate the Lifshitz point at 
$\kappa = 0.270 \pm 0.004$, $T_c = 3.7475 \pm 0.0005$, thus confirming
an estimation from a high temperature series expansion.\cite{Oit85}

The agreement of the independently estimated 
exponents $\alpha$, $\beta$ and $\gamma$  with the scaling 
relation $\alpha + 2 \beta+ \gamma = 2$, up to $\approx0.8\%$, 
illustrates the reliability of our data. 

We are now ready to discuss the scaling of the spin-spin correlation function
$C \left(\vec{r}_\perp, r_\parallel \right) = 
\left< s_{\vec{r}_\perp, r_\parallel}
\, s_{\vec{0},0} \right>$ and its scaling function $\Omega(v)$ as defined in
(\ref{Gl:DynS}). In $(d_{\perp}+1)$ dimensions, one has  
$\zeta= {2(d_\perp+ \theta)}/{\theta(2+\gamma/\beta)}$. 
For the $3D$ ANNNI model,
$\zeta =1.30\pm 0.05$, where the error follows from the errors in the values 
of the critical exponents.
In Figure 2 we show selected data for the function 
\begin{equation}
\Phi \left( u \right)=u^{-\zeta} \, \Omega(1/u)
\end{equation}
with $u = \sqrt{r_\perp\,}/{r_\parallel}$, as
computed by Monte Carlo simulations of a system with 
$200 \times 200 \times 100$ spins (assuming\cite{theta} $\theta=1/2$).
This permits a nice visual test of the data collapse and establishes scaling.

The small deviations (of order $\approx 2\%$) from 
the value $\theta=1/2$ obtained in recent field-theoretical 
calculations\cite{Die00} are not yet distinguishable\cite{theta} from the 
purely numerical errors in our data and the exponents derived from them. 
Therefore, for our purposes, namely the test of LSI, it is enough
to set $\theta=2/N=1/2$, leading to $N=4$ in the differential equation
(\ref{Gl:DiffGl}). In addition, the scaling form (\ref{Gl:DynS}) implies the
boundary condition $\Omega(v) \sim v^{-\zeta}$ for $v\to\infty$. For $N=4$, 
there are two independent solutions of eq.~(\ref{Gl:DiffGl}) satisfying this 
boundary condition\cite{Hen99,Hen97} and the scaling function becomes
\begin{equation} \label{Gl:OmeF}
\Omega(v) = b_0 F_0(v/(4 \alpha_1)^{1/4}) + b_1 v F_1(v/(4 \alpha_1)^{1/4})
\end{equation}
where\cite{Hen97} 
\begin{eqnarray}
F_0(x) &=& \frac{\Gamma(3/4)}{\Gamma(\zeta/4)}
\sum_{n=0}^{\infty} \frac{\Gamma(n/2+\zeta/4)}{n! \Gamma(n/2+3/4)} 
\left(-{x^2}\right)^{n}  \label{Gl:Funkt}\\
F_1(x) &=& \frac{\Gamma(3/2)}{\Gamma(\frac{1}{4}(\zeta+1))}
\sum_{n=0}^{\infty} \frac{\Gamma((n+1+\zeta)/4) s(n)}
{\Gamma(n/4+1)\Gamma(\frac{1}{2}(n+3))} (-{x})^n \nonumber
\end{eqnarray}
and $s(n)=\frac{1}{\sqrt{2}}(\cos\frac{n\pi}{4}+
\sin\frac{n\pi}{4})\cos\frac{n\pi}{4}$. 
Here $\alpha_1$ is the constant occuring in (\ref{Gl:DiffGl}) 
and $b_{0,1}$ are free parameters. Since
$b_0$ and $\alpha_1$ are merely scale factors, the functional form of the
scaling functions $\Omega(v)$ and 
$\Phi \left( u \right)=u^{-\zeta} \, \Omega(1/u)$ 
depends on the single {\em universal} parameter $p:= \alpha_1^{1/4} b_1/b_0$. 

To see this, consider the moments 
$M(n) := \int_{0}^{\infty} \!du \, u^n \,
\Phi(u)$.
For $\theta=1/2$ and taking into account (\ref{Gl:OmeF}), it
follows\cite{Ben84} that in the scaling region the moment ratios 
\begin{equation} \label{eqj}
J(\left\{m_i\right\};\left\{n_j\right\}) := 
\left. \prod\limits_{i=1}^k M(m_i) \right/ \prod\limits_{j=1}^k M(n_j)
\end{equation}
with $k\geq 2$ and $\sum_i m_i=\sum_j n_j$ are independent of 
$b_0$ and $\alpha_1$ and only depend on the functional form of $\Phi(u)$ as 
parametrized by $p$. Our Monte Carlo data for the spin-spin correlator will be 
consistent with LSI if the values of $p$ determined from several different 
ratios $J$ coincide.  

As we are not able to compute numerically the function $\Phi(u)$ for values
of $u$ below $u_0 \approx 0.22$ a direct analysis along the lines just 
sketched is not possible. Instead we have to consider 
the moments $\widetilde{M}(n) := \int_{u_0}^{\infty} \!du \, u^n \,
\Phi(u \, \alpha_1^{1/4}) = \alpha_1^{-(n+1)/4} 
\int_{w_0}^{\infty} \!dw \, w^n \, \Phi(w)$
with $w_0=u_0 \, \alpha_1^{1/4}$.
The moment ratios $\widetilde{J}(\left\{m_i\right\};\left\{n_j\right\})$
(defined in complete analogy with (\ref{eqj})) then depend on 
the scale factor $\alpha_1^{1/4}$ through $w_0$. 
The parameters $\alpha_1$ and $p$ are determined from the following scheme.
Choosing a suitable starting value for $\alpha_1$ 
we compute an approximative value for $p$ by 
comparing the values of the moment ratios $\widetilde{J}$ (obtained from the
full data set for $\Phi(u)$ as shown in the inset of Figure 2)
with the $p$-dependent expressions
coming from integrating the $\widetilde{M}(n)$ using 
the analytic form (\ref{Gl:OmeF},\ref{Gl:Funkt}).  
An improved value for $\alpha_1$ is then derived by comparing 
the values $\widetilde{M}(m)/\widetilde{M}(n)$ for arbitrary $m$ and $n$
obtained (i) from our numerical data and (ii) from the analytical 
expressions after inserting the value of $p$. The final values of
$\alpha_1$ are obtained by averaging over five different pairs $(m,n)$.
 
The values of $p$ and $\alpha_1$ 
determined from several distinct moment 
ratios are collected in Table II.
We obtain the mean values $p = -0.11(1)$ and $\alpha_1=33.2(8)$.
The consistency of the different determinations of the two 
parameters provides clear evidence in favour of the applicability of 
the hypothesis of local scale invariance to the Lifshitz point of the 
ANNNI model. 

Finally, $b_0 \approx 0.41$ is obtained by
adjusting the scale of $\Phi$. 
Inserting these values into the analytical
expression yields for $\Phi$ the full curve shown in the inset of Figure 2.
The agreement between our MC data and the theoretical result is remarkable.

Local scale invariance had been confirmed before at the Lifshitz point
of the ANNNS model. In that exactly solvable model, 
the Ising model spins in (\ref{Gl:annni})
are replaced by spherical model spins $s_{xyz}\in {\tt I\!R}$ together
with the usual spherical constraint.\cite{Sel92} 
At the Lifshitz point, one has 
$\theta=1/2$ and the exactly known spin-spin correlator\cite{Fra93} agrees
with the scaling form (\ref{Gl:OmeF}) for $b_1=0$.\cite{Hen97} Our finding
that LSI also appears to hold for the ANNNI model suggests
that the domain of validity of LSI should extend beyond the context 
of free field theory which underlies the ANNNS model. 
It appears plausible that LSI will also hold true for the Lifshitz 
points of the ANNNXY, ANNNH,\ldots models\cite{Sel92} which are intermediate 
between the ANNNI and the ANNNS model. 
Since the number of dimensions $d_{\perp}$ merely enters as a parameter, 
local scale invariance could also be tested along the lines of an
$\epsilon$-expansion.\cite{Die00} Finally, the consistency of other
correlators (e.g. energy-energy) with LSI should be tested.
We plan to come back to this elsewhere. It would be tempting
to see whether the powerful techniques of $2D$ conformal invariance\cite{Bel84} 
might be extended to the situation of anisotropic scaling realized at Lifshitz
points. This would lead to numerous physical 
applications\cite{Sel88,Yeo88,Neu98,Ska00,Vys92,Bat95,Sch98} and is under
investigation. 

In conclusion, the precise localisation of the $3D$ ANNNI model Lifshitz point 
and improved estimates of its critical exponents allowed for the first time
to determine reliably the scaling of the spin-spin correlator. 
Its functional form was found to agree with the prediction of local scale 
invariance. The confirmation of the applicability of local scale invariance
to this situation suggests a new symmetry principle
for the description of equilibrium systems with anisotropic scaling, especially
for systems with competing interactions at their Lifshitz points. 

Acknowledgements: We thank H.W. Diehl and M. Shpot for communicating their
two-loop results\cite{Die01} before publication and the CINES Montpellier
for providing substantial computer time (projet pmn2095). 


\begin{table}
\caption{Critical exponents at the Lifshitz point of the $3D$ ANNNI model,
as obtained from Monte Carlo simulations (MC), 
high-temperature series expansion (HT) and renormalized field theory (FT).
The numbers in brackets give the estimated error in the last digit(s).}
\begin{tabular}{|l|l|l|l|l|l|}
& $\alpha$ & $\beta$ & $\gamma$ & $(2-\alpha )/\gamma$ & $\beta/\gamma$ \\
\hline
MC [$\!\!$\onlinecite{Kas85}] & & $0.19(2)$ & $1.40(6)$ & & $0.14(2)$ \\
\hline
HT [$\!\!$\onlinecite{Mo91}]  & $0.20(15)$ & & $1.6(1)$ & $1.1(2)$ & \\
\hline
FT [$\!\!$\onlinecite{Die00}] & & & & 1.27& 0.134 \\
\hline
FT [$\!\!$\onlinecite{Die01}] & 0.160 & 0.220 & 1.399 & 1.315 & 0.157 \\ \hline
MC & $0.18(2)$ & $0.238(5)$ & $1.36(3)$ 
& $1.34(5)$ & $0.175(8)$\\
\end{tabular}
\end{table}

\begin{table}
\caption{Values of the parameters $p$ and $\alpha_1$ computed from different
moment ratios $\widetilde{J}(\left\{m_i\right\};\left\{n_j\right\})$, 
see text.}
\begin{tabular}{|l|l|l|l|}
$\left\{m_i\right\}$ & $\left\{n_j\right\}$ & $p$ & $\alpha_1$ \\
\hline
$\left\{0,-0.5\right\}$ & $\left\{-0.25,-0.25\right\}$ & $-0.102$ & 32.7\\
\hline
$\left\{-0.25,-0.75\right\}$ & $\left\{-0.5,-0.5\right\}$ & $-0.125$& 34.0\\
\hline
$\left\{0.2,-0.9\right\}$ & $\left\{0,-0.7\right\}$ & $-0.100$& 32.8\\
\hline
$\left\{0.2,-0.6,-0.8\right\}$&$\left\{-0.3,-0.4,-0.5\right\}$&$-0.102$& 32.8\\
\hline
$\left\{-0.1,-0.6,-0.7\right\}$&$\left\{-0.4,-0.5,-0.5\right\}$&$-0.117$&33.5\\
\end{tabular}
\end{table}

\begin{figure}
\centerline{\epsfxsize=2.75in\epsfbox
{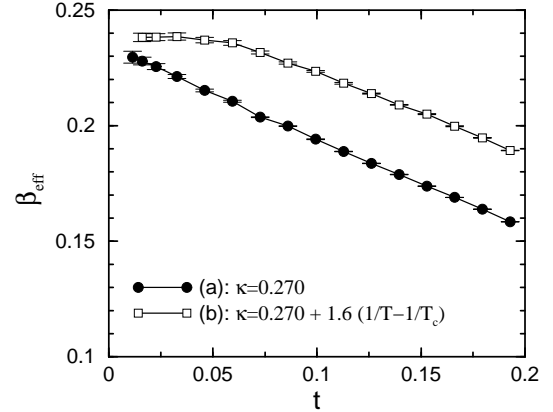}}
\caption{Effective exponent $\beta_{\rm eff}$ versus $t$ 
for two different trajectories in the
$(T,\kappa)$ space, see text. 
Error bars include the uncertainty in $T_c$: 
$T_c ( \kappa=0.270) = 3.7475 \pm 0.0005$.
\label{fig1}} \end{figure}

\begin{figure}
\centerline{\epsfxsize=2.75in\epsfbox
{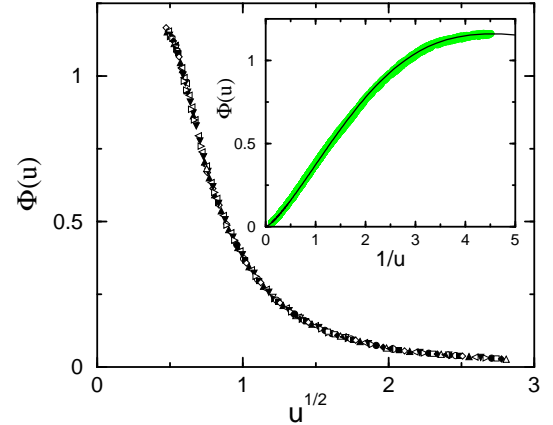}}
\caption{Scaling funtion $\Phi(u)$ (see text) versus 
$u^{1/2}=(\sqrt{r_\perp}/{r_\parallel})^{1/2}$ for $\kappa=0.270$ 
and $T=3.7475$. Selected Monte Carlo data for a system of 
$200 \times 200 \times 100$ spins are shown.\cite{theta}
The different symbols correspond to the values
of $r_\perp$. Inset: comparison of the full data set of $1.7 \cdot 10^4$
points for the scaling function
$\Phi(u)$ (gray points) with the analytical prediction 
eqs.~(\ref{Gl:OmeF},\ref{Gl:Funkt}) following from the assumption of LSI, with
$p=-0.11$, $\alpha_1=33.2$ and $b_0=0.41$ (full curve).
\label{fig2}} \end{figure}


\begin{references}
\bibitem{theta} Using $\theta=0.488$ as suggested in Ref. \onlinecite{Die00}
does not lead to a perceptible modification of the scaling plot.
\end{references}
\end{document}